\documentclass[twocolumn,times,tighten]{aastex62}

\newcommand{\eg}{\emph{e.g.}}
\newcommand{\ie}{\emph{i.e.}}
\newcommand{\cf}{\emph{cf.}}
\makeatletter

\newcommand{\Rmnum}[1]{\expandafter\@slowromancap\romannumeral #1@}
\makeatother

\newcommand{\corr}[1]{\textcolor{black}{#1}}
\newcommand{\corrtwo}[1]{\textcolor{black}{#1}}

\received{Apr 1, 2019}
\revised{May 1, 2019}
\accepted{May 1, 2019}
\submitjournal{ApJ}

\shorttitle{Multi-wavelength observations of homologous waves}
\shortauthors{Long et al.}

\begin{document}

\title{\corr{Quantifying the relationship between Moreton--Ramsey waves and ``EIT waves'' using} observations of 4 homologous \corr{wave events}}

\correspondingauthor{David M. Long}
\email{david.long@ucl.ac.uk}

\author[0000-0003-3137-0277]{David M.~Long}
\affil{UCL-Mullard Space Science Laboratory, Holmbury St. Mary, Dorking, Surrey, RH5 6NT, UK}

\author[0000-0002-8975-812X]{Jack~Jenkins}
\affil{UCL-Mullard Space Science Laboratory, Holmbury St. Mary, Dorking, Surrey, RH5 6NT, UK}

\author[0000-0001-7809-0067]{Gherardo~Valori}
\affil{UCL-Mullard Space Science Laboratory, Holmbury St. Mary, Dorking, Surrey, RH5 6NT, UK}

\begin{abstract}
Freely-propagating global waves in the solar atmosphere are commonly observed using Extreme UltraViolet passbands (EUV or ``EIT waves''), and less regularly in H-alpha (Moreton-Ramsey waves). Despite decades of research, joint observations of EUV and Moreton-Ramsey waves remain rare, complicating efforts to quantify the connection between these phenomena. We present observations of four homologous global waves originating from the same active region between 28--30~March~2014 and observed using both EUV and H-alpha data. Each global EUV wave was observed by the Solar Dynamics Observatory, with the associated Moreton-Ramsey waves identified using the Global Oscillations Network Group (GONG) network. All of the global waves exhibit high initial velocity \corr{(\eg, 842--1388~km~s$^{-1}$ in the 193~\AA\ passband)} and strong deceleration \corr{(\eg, -1437 -- -782~m~s$^{-2}$ in the 193~\AA\ passband) in each of the EUV passbands studied}, with the EUV wave kinematics exceeding those of the Moreton-Ramsey wave. The density compression ratio of each global wave was estimated using both differential emission measure and intensity variation techniques, with both indicating that the observed waves were weakly shocked with a \corr{fast magnetosonic} Mach number slightly greater than one. This suggests that, according to current models, the global coronal waves were not strong enough to produce Moreton-Ramsey waves, indicating an alternative explanation for these observations. Instead, we conclude that the evolution of the global waves was restricted by the surrounding coronal magnetic field, in each case producing a downward-angled wavefront propagating towards the north solar pole which perturbed the chromosphere and was observed as a Moreton-Ramsey wave.
\end{abstract}


\section{Introduction} \label{s:intro}

Globally-propagating waves in the solar atmosphere were first observed in the early 1960's by \citet{Moreton:1960a} and \citet{Moreton:1960b} using H$\alpha$ observations of the solar chromosphere. Although these Moreton--Ramsey waves were theorised by \citet{Uchida:1968} to be the result of a globally-propagating shock wave in the solar corona, this hypothesis could not be tested until the launch of the \emph{Solar and Heliospheric Observatory} \citep[SOHO;][]{Domingo:1995} with its full-Sun Extreme ultraviolet Imaging Telescope \citep[EIT;][]{Dela:1995} in 1995. The first observations of the so-called ``EIT waves'' reported by \citet{Dere:1997}, \citet{Moses:1997} and \citet{Thompson:1998} were therefore interpreted as the coronal counterpart of the chromospheric Moreton--Ramsey wave. 

\begin{deluxetable*}{cccccccccc}[t]
\tablecaption{\corr{List of events studied. Note that kinematic errors were calculated by fitting all of the data points shown in Figure~\ref{fig:stack_plot}} \label{tbl:events}}
\tablecolumns{8}
\tablewidth{0pt}
\tablehead{
\colhead{Flare peak time} & \colhead{Flare source} & \colhead{Flare} & \colhead{Angular}\tablenotemark{a} & \colhead{} & \multicolumn{5}{c}{Kinematics} \\
\colhead{UT} & \colhead{x, y (arcsec)} & \colhead{size} & \colhead{Extent} & \colhead{Quantity}\tablenotemark{b} & \colhead{171~\AA} & \colhead{193~\AA} & \colhead{211~\AA} & \colhead{304~\AA} & \colhead{H$\alpha$}}
\startdata
2014-03-28 & 339.3, 286.2 & M2.0 & 326$^{\circ}$--43$^{\circ}$ & vel. & 1044$\pm$169 & 1011$\pm$100 & 1025$\pm$181 & 996$\pm$211 & 566$\pm$66 \\ 
19:18:00 & & & & acc. & -834$\pm$240 & -856$\pm$200 & -876$\pm$253 & -839$\pm$355 & -589$\pm$384 \\
2014-03-28 & 369.9, 284.6 & M2.6 & 320$^{\circ}$--19$^{\circ}$ & vel. & 1498$\pm$511 & 1388$\pm$175 & 1404$\pm$96 & 1036$\pm$143 & 782$\pm$57 \\  
23:51:00 & & & & acc. & -1662$\pm$1130 & -1437$\pm$256 & -1458$\pm$143 & -1050$\pm$481 & -1329$\pm$374 \\
2014-03-29 & 501.5, 275.9 & X1.0 & 320$^{\circ}$--39$^{\circ}$ & vel. & 1200$\pm$140 & 1231$\pm$287 & 1215$\pm$216 & 974$\pm$158 & 667$\pm$26 \\  
17:48:00 & & & & acc. & -1117$\pm$178 & -1158$\pm$332 & -1148$\pm$255 & -885$\pm$369 & -307$\pm$93 \\
2014-03-30 & 650.6, 213.8 & N/A\tablenotemark{c} & 321$^{\circ}$--29$^{\circ}$ & vel. & 834$\pm$81 & 842$\pm$149 & 934$\pm$182 & 908$\pm$200 & 444$\pm$78 \\  
11:48:00 & & & & acc. & -803$\pm$154 & -782$\pm$300 & -909$\pm$379 & -976$\pm$487 & 45$\pm$542 \\
\enddata
\tablenotetext{a}{Measured clockwise from solar north}
\tablenotetext{b}{Units are km~s$^{-1}$ for velocity and m~s$^{-2}$ for acceleration}
\tablenotetext{c}{Due to a data gap, the peak X-ray flux was not measured by GOES for this flare.}
\end{deluxetable*}

Despite this initial assumption, discrepancies in the kinematics, morphology and behaviour of these ``EIT waves'' led to the development of a series of alternative theories contained within two main branches to explain the phenomenon. On the wave branch, in addition to the fast-mode \corr{magnetoacoustic wave} interpretation originally used, ``EIT waves'' were alternatively interpreted as slow-mode magnetoacoustic waves \citep[][]{Wang:2009}, magnetohydrodynamic solitons \citep[\cf][]{Wills-Davey:2007} or large-amplitude waves or shocks \citep[\cf][]{Vrsnak:2008}. The second main branch interpreted ``EIT waves'' as a signature of magnetic field restructuring during the eruption of a coronal mass ejection (CME), identifying the propagating bright front as being alternatively due to Joule heating at the boundary of a current shell \citep{Delannee:2007,Delannee:2008}, continuous reconnection with nearby small-scale quiet-Sun loops \citep{Attrill:2007} or stretching of magnetic field lines during the eruption of the CME \citep{Chen:2002}. A full overview of the different theories proposed to explain the ``EIT wave'' phenomenon may be found in the recent review by \citet{Warmuth:2015}. More recently, the advent of high temporal and spatial observations across multiple passbands provided by the \emph{Solar Dynamics Observatory} \citep[SDO;][]{Pesnell:2012} spacecraft and the multiple points of view provided by the \emph{Solar Terrestrial Relations Observatory} \citep[STEREO;][]{Kaiser:2008} spacecraft have led to a consensus within the community that ``EIT waves'' are large-amplitude waves or shocks \citep[see][for more details]{Long:2017a}. Note that we shall refer to ``EIT waves'' as EUV \corr{waves} for the remainder of this manuscript \corr{to highlight the fact that they are observed using instruments other than SOHO/EIT and to ensure consistency in terminology with the Moreton--Ramsey waves observed using the H-alpha passband}.

Part of the issue with regard to the uncertainty surrounding the physical interpretation of EUV \corr{waves} and their relationship to Moreton--Ramsey waves is due to the paucity of \corr{simultaneous} observations of both phenomena. Although relatively rare, \citet{Nitta:2013} identified 171 EUV \corr{waves} between April~2010 and January~2013 \citep[extended to 362 EUV \corr{waves} identified between April~2010 and August~2016 by][]{Long:2017b}, primarily due to the high synoptic cadence of the Atmospheric Imaging Assembly \citep[AIA;][]{Lemen:2012} onboard the SDO spacecraft. However, observations of Moreton--Ramsey waves remain frustratingly rare, despite the worldwide coverage of the Global Oscillations Network Group (GONG) telescope network. Since the launch of SDO in 2010, fewer than 5 Moreton--Ramsey waves have been identified and analysed in detail. \citet{Asai:2012} reported on a joint Moreton--Ramsey and EUV wave from 9~August~2011 observed using the Solar Magnetic Activity Research Telescope at Hida Observatory, while \citet{Francile:2016} reported on a joint Moreton--Ramsey and EUV wave from 29~March~2014 observed using the H-Alpha Solar Telescope for Argentina (HASTA). Prior to the launch of SDO, several authors reported on high-cadence observations of Moreton--Ramsey waves, usually with one or two observations of an associated EUV \corrtwo{wave} on a similar kinematic trajectory \citep[\eg][]{Warmuth:2004a,Warmuth:2004b,Veronig:2006,Bala:2010,Muhr:2010}. 

This discrepancy between the number of global EUV \corr{waves} and Moreton--Ramsey waves remains a source of \corr{ongoing} investigation. Independent simulations performed by both \citet{Vrsnak:2016} and \citet{Krause:2018} suggest that a strong overexpansion of an erupting flux rope during the initial stages of a solar eruption is required to produce a propagating shock wave strong enough to perturb the chromosphere and be observed as a Moreton--Ramsey wave. \citet{Vrsnak:2016} suggest that this is a result of the pressure jump associated with the passage of the coronal shock, which produces a downward-propagating quasi-longitudinal MHD shock (well approximated in their model by a switch-on MHD shock). If sufficiently strong, this downward propagating shock can produce the observed Moreton--Ramsey wave. For weaker events, or if the lateral overexpansion of the flux rope is not sufficiently strong, \citet{Vrsnak:2016} suggest that a Moreton--Ramsey wave could still be produced if the eruption is highly asymmetric. Although strong overexpansion of the CME bubble has been previously observed to drive global EUV waves \citep[\cf][]{Patsourakos:2010,Veronig:2018}, it has not yet been observed to drive Moreton--Ramsey waves. \corr{However, Moreton--Ramsey waves have traditionally been observed to be arc-shaped and therefore anisotropic \citep[\cf][]{Warmuth:2015}, suggesting that they may be produced by a highly-asymmetric eruption. The four events presented here provide the opportunity to test this hypothesis for the relationship between the EUV and Moreton--Ramsey waves, with the EUV waves well observed by SDO/AIA and the Moreton--Ramsey waves well observed by the GONG network (albeit using the H-alpha line core rather than the wings).}

\begin{figure*}[!ht]
\centering
\includegraphics[width = 1\textwidth, trim=0 0 0 0,clip=]{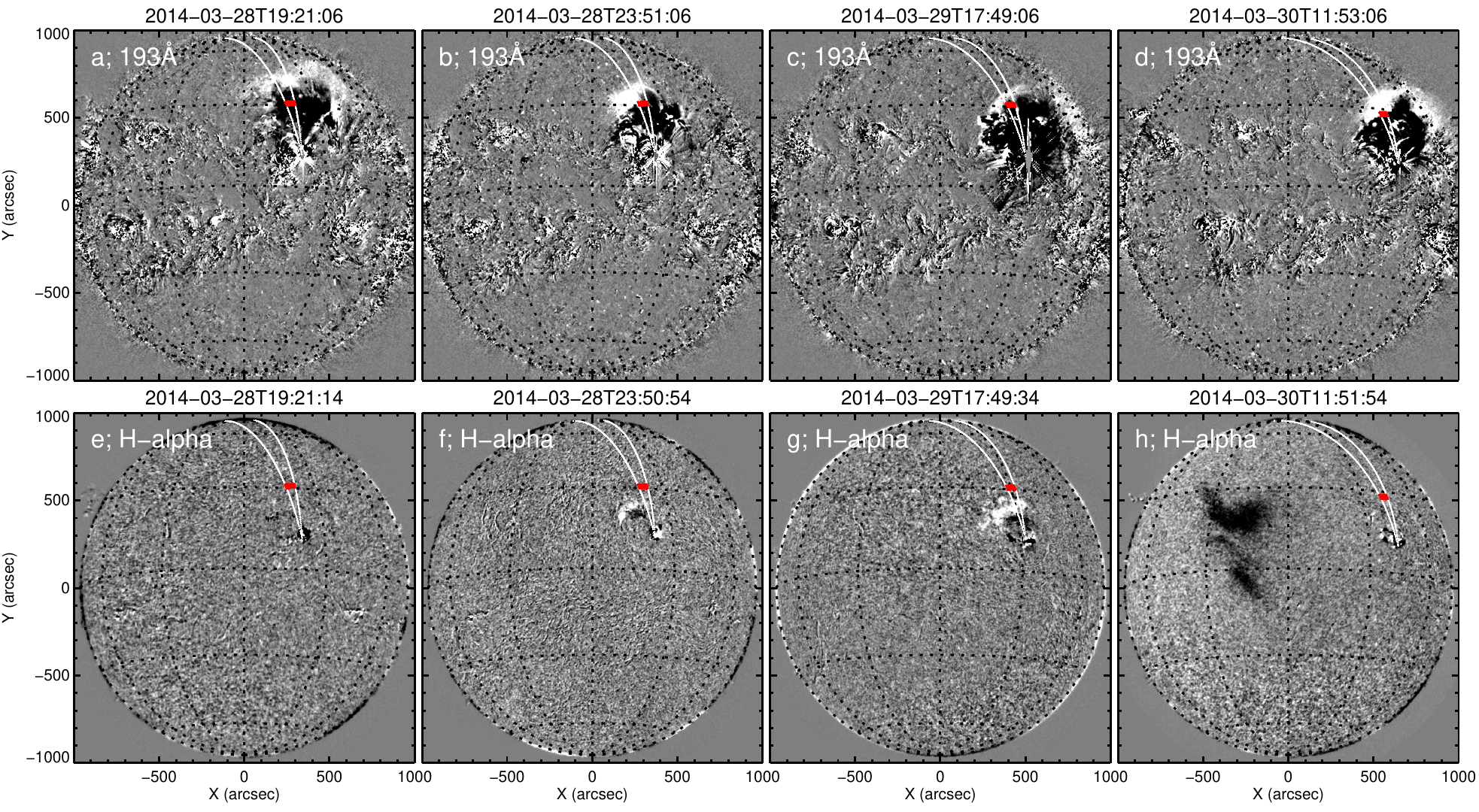}
\caption{The four global wave events studied in this work. The top row shows the events observed using the 193~\AA\ passband on SDO/AIA, while the bottom row shows the same events as observed using the GONG H$\alpha$ passband. Each panel shows a running difference image, with the leading image taken at the time shown and the following image 120~s earlier. The red region shows the location used to estimate the variation in density and temperature in Section~\ref{ss:diag}.}
\label{fig:context}
\end{figure*}


In this paper, we \corr{use} observations of four homologous global Moreton--Ramsey and EUV waves which erupted from the same active region from 28--30~March~2014 \corr{to quantify the relationship between these phenomena}. The observations are presented in Section~\ref{s:obs}, with the analysis and results presented in Section~\ref{s:res}. These results are then discussed in Section~\ref{s:disc} before some conclusions are drawn in Section~\ref{s:conc}.

\section{Observations and Data Analysis} \label{s:obs}

The four global wave events studied here are outlined in Table~\ref{tbl:events} and erupted from NOAA Active Region (AR) 12017 over the course of three days from 28--30~March~2014. The different events are shown in Figure~\ref{fig:context}, with the top row showing running-difference images of the events in the 193~\AA\ passband from SDO/AIA, while the bottom row shows the corresponding running-difference H-alpha observations for each event. Each event was associated with a GOES X-ray flare, a coronal mass ejection (CME) identified by the LASCO CDAW catalogue\footnote{\url{https://cdaw.gsfc.nasa.gov/CME_list/UNIVERSAL/2014_03/univ2014_03.html}}\corr{, and a Type~\corrtwo{\Rmnum{2}} radio burst as measured by NOAA/SWPC\footnote{\corrtwo{see, \eg, \url{https://www.solarmonitor.org/data/2014/03/29/meta/noaa_events_raw_20140329.txt}}}}. Table~\ref{tbl:events} shows that with the exception of the flare on 2014-03-30, which was not measured due to a data gap, all of the flares were quite large, ranging from M2.0--X1.0. 

The events were all well observed by the SDO/AIA, with the global EUV waves identifiable in each of the EUV passbands studied. However, only the 171, 193, 211, and 304~\AA\ passbands were used for this analysis as the signal-to-noise was too low in the 94, 131, and 335~\AA\ passbands. In each case, the data were \corr{reduced} and aligned using the standard SolarSoftWare routines \citep{Freeland:1998}. Although the Coronal Pulse Identification and Tracking Algorithm \citep[CorPITA;][]{Long:2014} was initially applied to the data from each SDO/AIA passband to identify and characterise the global \corr{EUV} wave, it was originally optimised for the 211~\AA\ passband and as a result could not accurately and consistently track the global EUV wave observed in the other passbands. This is due to the significant differences in intensity variation of both the \corr{EUV wave} and individual features in the surrounding corona in different passbands, all of which can result in the algorithm being unable to consistently identify and track the propagating bright feature. This is a known issue with CorPITA, and one that the code is currently being updated to try and overcome.

\begin{figure*}[!ht]
\centering
\includegraphics[width = 1\textwidth, trim=0 0 0 0,clip=]{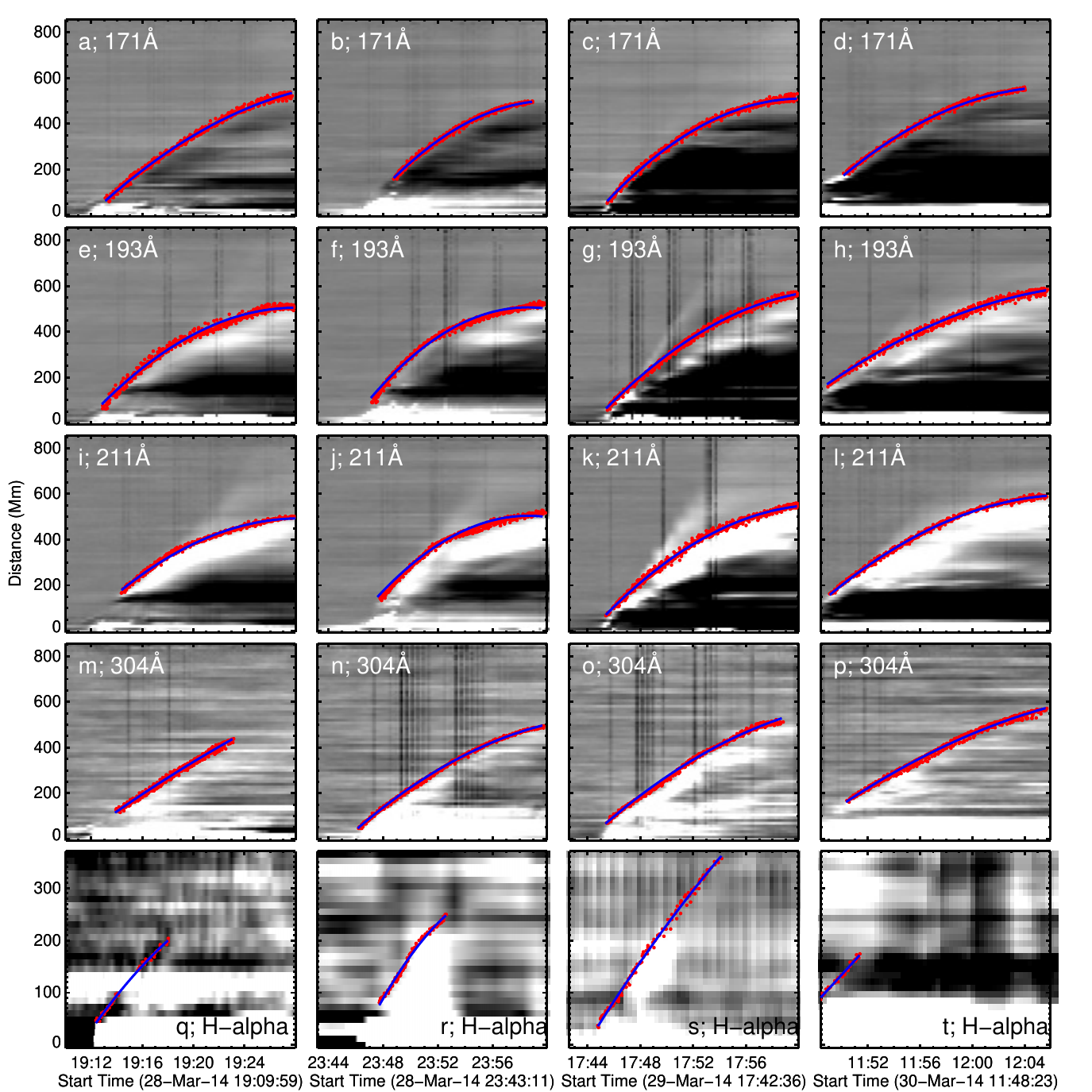}
\caption{Stack plots showing the variation in base difference intensity along an arc sector directed towards solar north in the 171~\AA, 193~\AA, 211~\AA, 304~\AA\ and H$\alpha$ passbands for each of the four events studied. \corr{In each case, the data-points indicating the manual selection of the wavefronts are shown in red, the} best-fit kinematics \corr{are} indicated by the solid blue lines, \corr{and} the mean values across all arc sectors \corr{are} given in Table~\ref{tbl:events}. Note that while the global \corr{waves} are easily identifiable in the 4 EUV passbands, they are much fainter in the H$\alpha$ passband, \corr{with the waves in panels q \& t dark instead of bright. We believe that this is most likely due to the dynamics of the chromospheric plasma contributing to the H-alpha line.} Although two intensity increases can be observed for each event in the 193~\AA\ and 211~\AA\ passbands, the bright, slow feature is identified here as the global \corr{wave}, whereas the faint, fast feature departing tangentially from the blue fit is the shock associated with the erupting coronal mass ejection previously identified by \citet{Francile:2016}. \corr{While the fits to the 193~\AA\ and 211~\AA\ data in panels f \& j turn over before the end time, this is unphysical and indicates that the fits in these cases have significant uncertainties. \corrtwo{The increased uncertainty in these particular fits} is accounted for in Table~\ref{tbl:events}, which gives the mean initial velocity and acceleration for all arcs across the full angular extent of the EUV waves.}}
\label{fig:stack_plot}
\end{figure*}

Instead, the intensity profiles derived by CorPITA were combined to produce a series of distance-time stack plots which were then used to manually identify the leading edge of the EUV wave with time for each passband studied. \corr{Each panel in Figure~\ref{fig:context} shows} the white arc sector\corr{s} used to make the stack plots shown in Figure~\ref{fig:stack_plot}. This approach was used to ensure consistency between passbands and events and enable the systematic calculation of errors when estimating the kinematics shown in Table~\ref{tbl:events}. The \corr{EUV waves were then manually identified by selecting} the leading edge of the bright feature in each arc sector for each passband and each event \corr{using 100 data-points}. This process was repeated 5 times for each arc sector to minimise user bias and ensure an accurate identification of the front. The kinematics were then estimated for each arc sector using a quadratic model \corr{applied to all 500 data points}, with the mean initial velocity and acceleration values for each passband listed in Table~\ref{tbl:events}. 

In addition to the EUV observations obtained by SDO/AIA, all four of the events were also observed in the H-alpha passband by the Global Oscillation Network Group (GONG) network of H-alpha telescopes. The GONG network consists of a series of 6 telescopes located around the world at Learmonth Solar Observatory in Western Australia, Big Bear Solar Observatory in California, USA, the High Altitude Observatory on Mauna Loa in Hawaii, USA, Udaipur Solar Observatory in India, the Observatorio del Teide in the Canary Islands and the Cerro Tololo Interamerican Observatory in Chile. Each of the different Moreton--Ramsey wave events were observed by a combination of telescopes, complicating the resulting analysis due to variations in seeing conditions. Although the Moreton--Ramsey waves could be visually identified using movies, they are harder to identify in individual images and indeed in the stack plots (as shown in Figure~\ref{fig:stack_plot}) as a result of this discrepancy between observatories. This was overcome by treating the data from each observatory independently and then combining the processed images to identify the Moreton--Ramsey waves.

\begin{figure*}[!ht]
\centering
\includegraphics[width = 0.97\textwidth, trim=0 0 0 0,clip=]{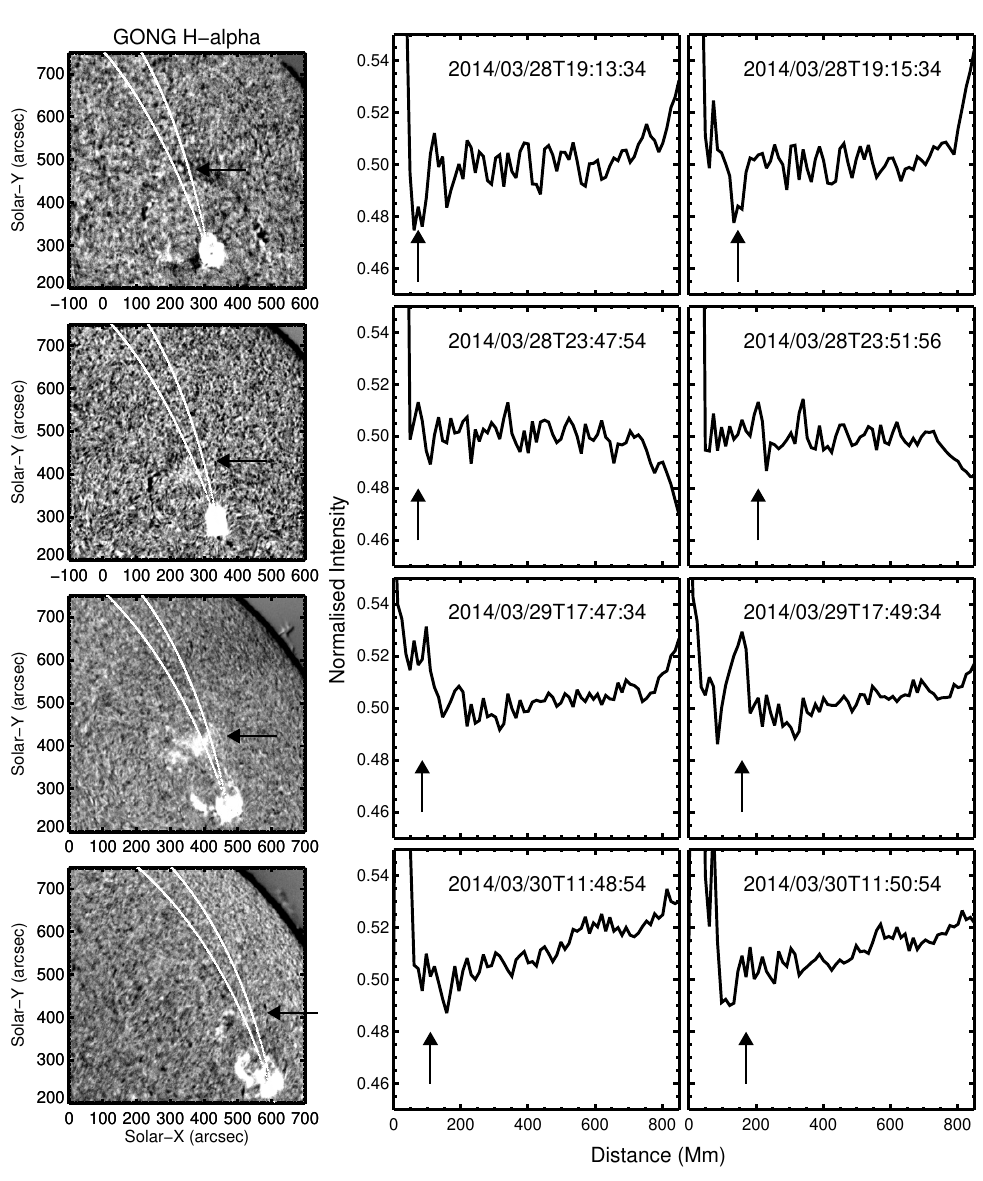}
\caption{Intensity images (left column) and intensity plots (centre and right columns) showing the Moreton--Ramsey waves associated with each of the four events studied here. Arrows in \corr{each column} indicate the positions of the identified Moreton--Ramsey waves. Note that the Moreton--Ramsey waves manifest as an increase in intensity for the bottom three events but as a decrease in intensity for the top event. \corr{Although this is most likely due to the H-alpha line being shifted due to the down-up swing typically associated with Moreton-Ramsey wave observations, it is not possible to confirm this hypothesis without observations from the H-alpha line wings, which are not available for these events.}}
\label{fig:halpha}
\end{figure*}

The data from each GONG observatory were aligned by first aligning the single image with the best seeing conditions \corr{identified by-eye} closest in time to the flare to the relevant image from the 304~\AA\ passband observed by SDO/AIA. Angular variations between different GONG observatories were then accounted for by deprojecting the subsequent images to polar coordinates and using different features to ensure accurate cross-correlation between all of the images. Each image for a given GONG observatory were then normalised with respect to the first image to ensure a consistent intensity range and to counteract variations in intensity due to atmospheric seeing. As a result of this procedure of first aligning a GONG image to an AIA image and then aligning all subsequent GONG images to the initial aligned image, the same arc sectors could be used to estimate kinematics from all of the EUV and GONG images. A base difference approach (\ie, subtracting the first image from all subsequent images for each observatory) was then used to identify the Moreton--Ramsey wave in both stack plots (\cf\ Figure~\ref{fig:stack_plot}) and individual intensity profiles (\cf\ Figure~\ref{fig:halpha}). 

\section{Results} \label{s:res}



\subsection{Pulse Characteristics}

The events described here displayed similar characteristic behaviour, propagating anisotropically away from the erupting active region towards solar north with a comparable angular extent (as shown in Figure~\ref{fig:context} and outlined in Table~\ref{tbl:events}). However, in contrast to the short time periods over which the events previously studied by \citet{Kienreich:2011,Kienreich:2013} and \citet{Zheng:2012} occurred ($\approx$10~hours, $\approx$12~hours and $\approx$3~hours respectively), the four events studied here occurred over the course of $\approx$42~hours between 19:00~UT on 2014-March-28 and 13:00~UT on 2014-March-30, suggesting some long-lasting property of the environment where the wave propagation took place. In addition, although $\approx$20 flares erupted from this active region over the same time period\footnote{\corrtwo{\cf\ \url{https://www.solarmonitor.org/data/2014/03/29/meta/noaa_events_raw_20140329.txt}}}, only the four events studied here were associated with global waves, suggesting a common initiation process. All four events were also seen to originate from the same part of the active region on its northern side, suggesting a homologous nature in the triggering mechanism and property of the erupting structure. The four events can therefore be considered to share comparable geometrical properties and originate and behave in a similar manner \citep[\cf][]{Kienreich:2011,Kienreich:2013,Zheng:2012}.

Note that following the work of \citet{Francile:2016}, the lower, more intense feature observed in the second and third rows of Figure~\ref{fig:stack_plot} (corresponding to the on-disk global wave rather than the projection of the erupting CME) was used to estimate the kinematics of the global wave in the 193~\AA\ and 211~\AA\ passbands. Although the global waves were quite clear in the different EUV passbands with minimal processing (as seen in Figure~\ref{fig:context}), the associated Moreton--Ramsey waves were much more difficult to identify in the H-alpha passbands. However, following the processing approach outlined in Section~\ref{s:obs}, it was possible to identify a leading edge in each case which could then be fitted to estimate the kinematics.

Each of the \corr{EUV wave} events studied here can be seen from Table~\ref{tbl:events} to \corr{have had a high initial velocity} with strong negative acceleration. Although no clear correlation can be observed between the flare intensity and the kinematics of the EUV wave, this is not unexpected as it was previously discussed by \citet{Long:2017b}. In fact, the average wave speed was higher for event 2 (with its M2.6 flare) than for event 3 (with its X1.0 flare), despite the smaller time gap between events 1 and 2 compared to events 2 and 3. \corr{The average EUV wave initial velocities (particularly in the commonly used 193~\AA\ and 211~\AA\ passbands) are comparable with other EUV wave events observed using SDO \citep[\cf][]{Nitta:2013,Long:2017b}. Although the kinematics estimated using the 304~\AA\ passband are lower than the other three EUV passbands, this is consistent with previous observations \citep[\cf][]{Long:2008}.}

The Moreton-Ramsey waves studied here were first identified `by-eye' using images taken in the line core of the H-alpha line by the GONG observatories. Although apparent in moving images due the ability of the human eye to detect motion, the Moreton-Ramsey waves were much more difficult to identify in individual images. Figure~\ref{fig:halpha} shows a combination of still images (left column) with arc sectors along which the intensity could be taken to try and identify the \corr{waves} (centre and right columns). The arrows in each of the plots in the centre and right columns indicate the Moreton-Ramsey \corr{wave}, enabling its temporal evolution to be tracked. \corrtwo{As the Moreton-Ramsey waves were quite difficult to identify for each event}, the intensity variation \corrtwo{in the intensity profile plots} corresponding to the Moreton-Ramsey wave was identified by \corrtwo{moving back and forth between} consecutive images and intensity profiles to determine a moving feature. \corrtwo{The evolution of the Moreton-Ramsey wave was shown in the stack plot shown in Figure~\ref{fig:stack_plot} by scaling the stack plot to highlight a sloping feature (indicative of a propagating front) and then manually identifying the leading edge of that sloping feature using a point-and-click approach (producing the points shown in panels q--t of Figure~\ref{fig:stack_plot}).} This \corrtwo{approach} enabled the features identified by the arrows in Figure~\ref{fig:halpha} to be determined and matched to the features identified in the bottom row of Figure~\ref{fig:stack_plot}.

These intensity plots at each moment in time could then be combined to produce the stack plots shown in the bottom row of Figure~\ref{fig:stack_plot}, enabling a determination of the temporal evolution of the Moreton-Ramsey waves for each event. \corr{As with the estimation of the kinematics for the EUV waves, the Moreton-Ramsey wave kinematics were estimated by repeating the identification of the leading edge of the propagating front 5 times and fitting the resulting cloud of data points with a quadratic function to estimate the initial velocity and acceleration.} In contrast to previously studied events, the Moreton--Ramsey waves identified here were found to be much slower than the associated global EUV waves. However, this may be a function of the lower cadence H-alpha observations (20--60~s) compared to the EUV observations (12~s) \citep[\cf][]{Byrne:2013}, and the difficulty in identifying the Moreton--Ramsey waves in the H-alpha observations. Moreton--Ramsey waves are thought to be the result of a coronal wave pressing down on the chromosphere, making them easier to observe using the wings of H-alpha due to a characteristic down-up swing in Doppler velocity. However, the downward force exerted by the coronal wave would need to be particularly large to be observed in line core images. This is not always the case, making it difficult to identify Moreton-Ramsey waves in H-alpha line core images.

\subsection{Coronal Plasma Variation}\label{ss:diag}

The strength of the global EUV \corr{wave} shock was examined by tracking the evolution of the intensity of the 193~\AA\ passband within the region highlighted in Figure~\ref{fig:context} \corr{\citep[\cf][]{Long:2015}. Assuming that the shock wave observed here is propagating perpendicular to the direction of the magnetic field \citep[consistent with previous work, \cf][]{Vrsnak:2002,Zhukov:2011}, t}he magnetosonic Mach number of the \corr{wave} $M_{ms}$ can be estimated using,
\begin{equation}
    M_{ms} = \sqrt{\frac{X(X + 5 + 5\beta)}{(4 - X)(2 + 5\beta/3)}},\label{eqn:mach}
\end{equation}
where X is the density compression ratio, defined as $X = n/n_0$, and $\beta$ is the plasma-$\beta$ \citep[here assumed to be 0.1 after][]{Muhr:2011}. The density compression ratio $X$ was estimated using two different approaches; indirectly from the intensity variation in the EUV images and directly from density estimates obtained using a differential emission measure approach. 

As discussed by \citet{Zhukov:2011}, the density compression ratio $X$ can be related to the variation in intensity of the 193~\AA\ passband via the approximation,
\begin{equation}
    \frac{n}{n_0} = \sqrt{\frac{I}{I_0}}, \label{eqn:dens}
\end{equation}
where $I_0$ and $n_0$ are the intensity and density respectively prior to the passage of the global wavefront. Although EUV intensity is a function of both temperature and density, this approach assumes that the change in temperature is small, enabling an estimate to be made of the change in density. 

The cadence and number of passbands provided by SDO/AIA have led to the development of a variety of techniques for estimating the differential emission measure (DEM) of the coronal plasma observed by SDO/AIA \citep[\eg,][]{Hannah:2013,Cheung:2015}. The DEM, $\phi(T)$, is defined as,
\begin{equation}
\phi(T)=n_{e}^{2}(T)\frac{dh}{dT},
\end{equation}
where $n_{e}$ is the electron density, and enables an alternative, direct estimation of how the density and temperature of the corona vary during the passage of the global EUV \corr{wave}\corrtwo{.} This can then be used to confirm the observations made using the intensity variation described in Equation~\ref{eqn:dens}. The regularised inversion technique developed by \citet{Hannah:2013} was used here to examine the variation in DEM-weighted average density and temperature of the region highlighted in red in Figure~\ref{fig:context} using the approach of \citet{Vann:2015}. Following \citet{Cheng:2012}, the DEM-averaged temperature and density can be defined as,
\begin{equation}
T = \frac{\int \phi(T) T dT}{\int \phi(T)dT},
\end{equation}
and,
\begin{equation}
n = \sqrt{\frac{\int\phi(T)dT}{h}},
\end{equation}
respectively, where $h$ is the scale height \citep[taken here as 90~Mm following observations by \eg,][]{Patsourakos:2009}. This enabled an estimate to be made of the variation in both average temperature and density in the region of interest using the DEM derived directly from the AIA observations. 

\begin{figure*}[!ht]
\centering
\includegraphics[width = 1\textwidth, trim=0 0 0 0,clip=]{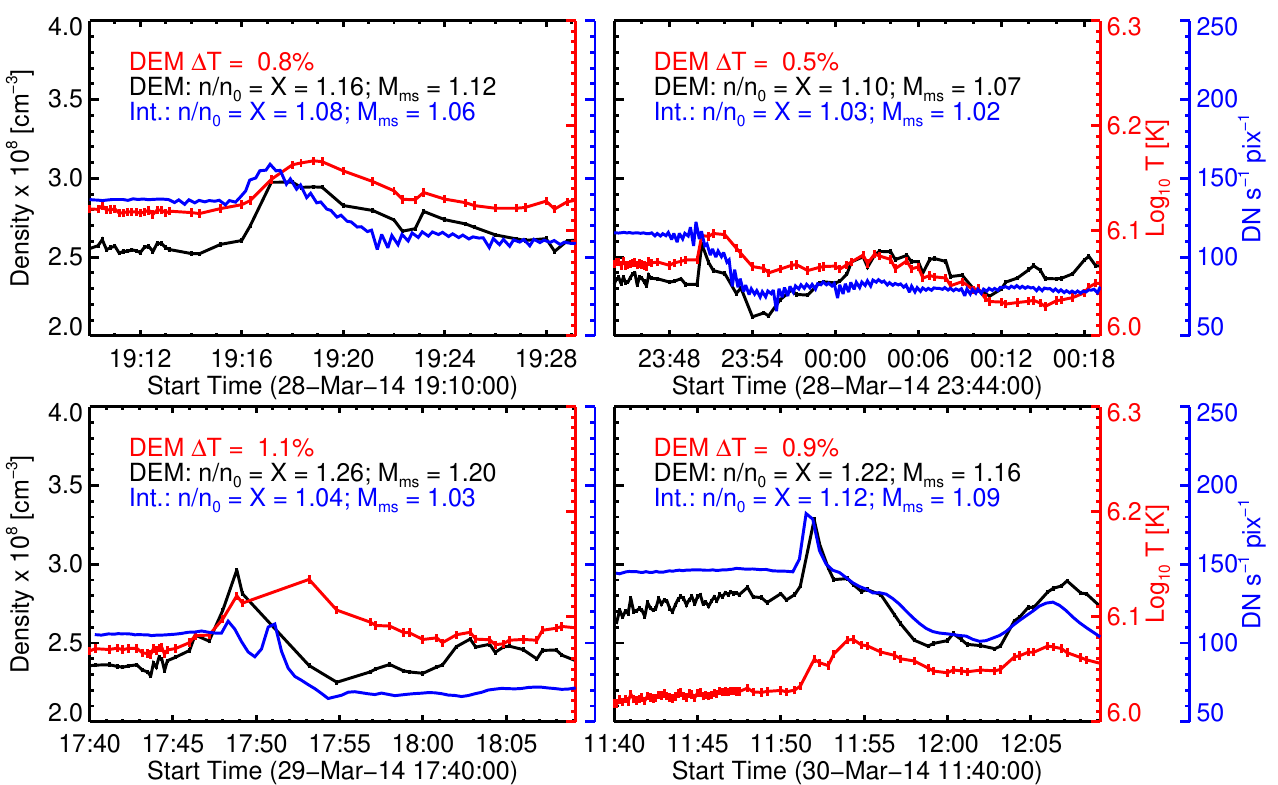}
\caption{The variation in AIA 193~\AA\ intensity (blue), and both density (black) and temperature (red) estimated using the DEM technique of \citet{Hannah:2013} with time for each of the events studied here. In each case the measurements were made at the location highlighted in red in Figure~\ref{fig:context}. The legend lists the percentage increase in density and temperature, the ratio of peak image intensity to the image intensity prior to the passage of the global \corr{wave}, and the magnetosonic Mach number estimated using Equation~\ref{eqn:mach}.}
\label{fig:int_vary}
\end{figure*}

The variation in image intensity (blue), DEM-derived density (black) and DEM-derived temperature (red) are shown for each event in Figure~\ref{fig:int_vary}. It is clear that each parameter exhibits an increase as a result of the passage of the global \corr{wave}. In each case, the intensity of the 193~\AA\ passband increases first, followed by the average density and finally the average temperature. Although the DEM-estimated average temperature exhibits an increase due to the passage of the \corr{wave}, in each case the percentage increase is quite small ($\lesssim$1.1~\%). This \corr{is consistent with the work discussed by \citet{Zhukov:2011}, and the weak nature of the shocks presented here, and} indicates that \corr{estimating} the change in image intensity as being due to the change in density (as in Equation~\ref{eqn:dens}) \corr{is a valid approximation} in this case. The variation in density and temperature are also consistent with the work of \citet{Vann:2015}, although the percentage increases (decreases) in density (temperature) are much larger (smaller) than previously found. 

Figure~\ref{fig:int_vary} also shows the magnetosonic Mach numbers using the shock compression ratio estimated by both intensity variation and DEM approaches. The Mach number estimated using the change in image intensity is consistently lower than that estimated using the DEM approach. This is most likely due to the fact that the DEM returns an estimate of the plasma distribution as a function of temperature integrated along the line of sight. All of the wavepulses studied here are propagating towards the north pole from an active region in the northern hemisphere, resulting in an increased amount of plasma along the line of sight contributing to the DEM solution. \corr{This increase in plasma along the line of sight as a function of latitudinal wave position results in an increased density value estimated using the DEM technique described here. \corrtwo{This is a consequence of the depth of the line of sight plasma contribution greatly exceeding the scale height assumed to estimate the density}.} However, all of the estimated Mach numbers are consistent with the previous events studied by \citet{Long:2014} and \citet{Muhr:2011}. The small estimated Mach numbers in each case are consistent with a weakly shocked global \corr{wave} \citep[\cf][]{Long:2017a}.

\section{Discussion}\label{s:disc}

\begin{figure*}[!ht]
\centering
\includegraphics[width = 1\textwidth, trim=0 0 0 0,clip=]{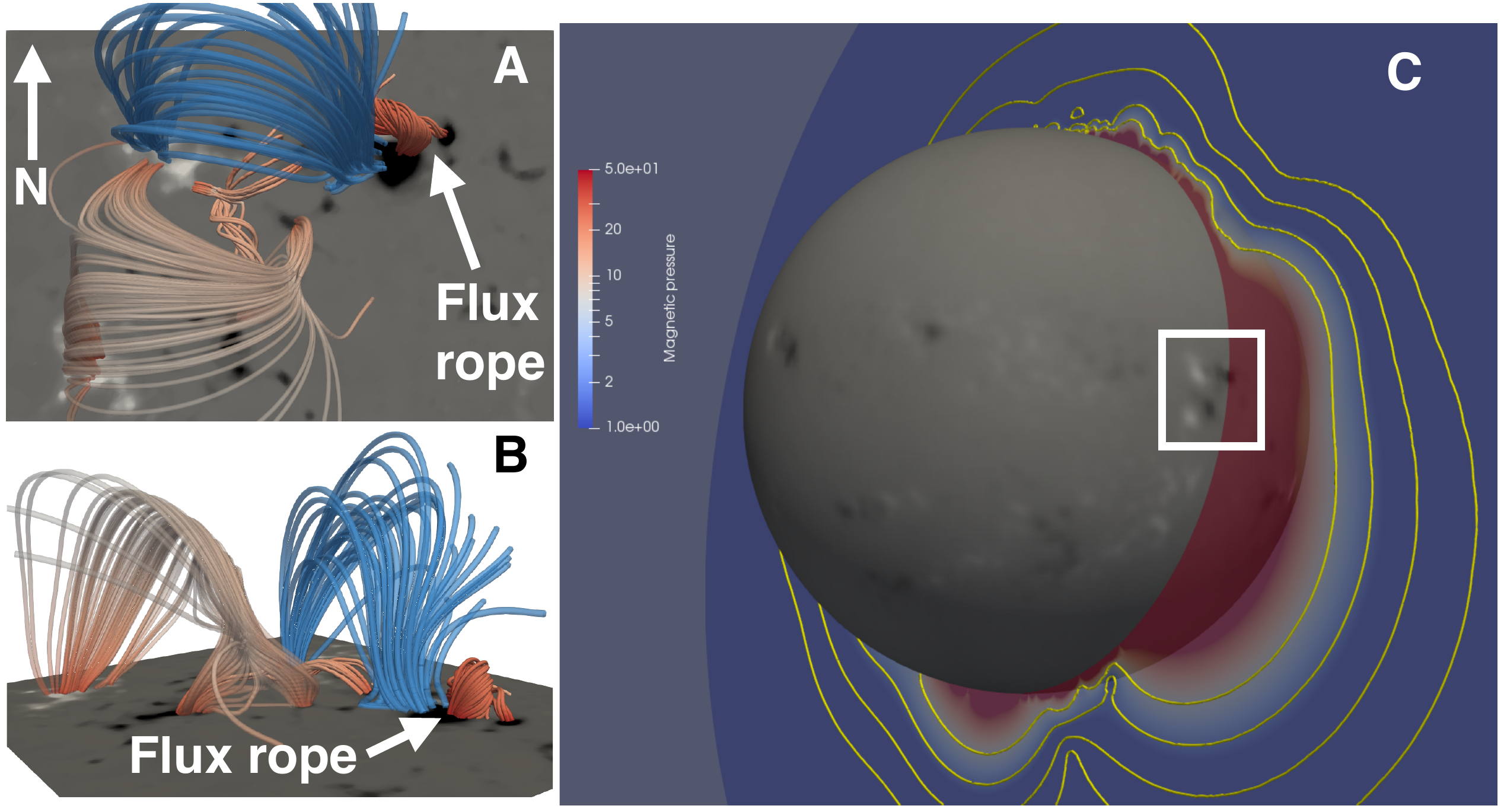}
\caption{A zoom-in of the magnetic field of AR12017 on 29-March-2014 using a nonlinear extrapolation from above (panel~a) and the side (panel~b), \corr{with the red/blue/white field lines indicating different domains of magnetic field.} Panel~c shows the calculated \corrtwo{magnetic pressure in arbitrary units of $\simeq3.5\times10^{-10}$ dyne/cm$^2$} along the arc sector shown in Figure~\ref{fig:context} from a PFSS extrapolation \corr{at 12:04~UT on 28-March-2014}. \corr{The white square in panel~c indicates the region of the magnetogram shown in panels~a \& b, while the yellow contours indicate lines of constant magnetic pressure.} Contrast the significant increase in magnetic pressure to the south of the erupting active region with the drop in magnetic pressure towards the north pole due to the increased magnetic complexity of the active region.}
\label{fig:b_field}
\end{figure*}

The four events presented here offer a rare opportunity to investigate and quantify how a global coronal wave pulse can drive a Moreton--Ramsey wave in H-alpha observations. The four events originate from the same active region over the course of $\approx$42 hours and can be considered to be homologous, with comparable geometry, kinematics and evolutionary behaviour. Although quite fast, none of the four events was particularly strong, with Mach numbers of $\approx1.03-1.20$ (depending on technique used to estimate them). This is comparable to previously studied events \corr{\citep[\eg][]{Long:2015}}, none of which produced Moreton--Ramsey waves, suggesting an additional requirement is needed in order to produce a response in the H-alpha line core observations discussed here. 

Simulations performed by \citet{Vrsnak:2016} suggested that a chromospheric response in H-alpha observations could be induced by a sufficiently strong density perturbation propagating in the corona. In this scenario, the density perturbation is produced by the rapid overexpansion of an impulsive erupting flux rope, which acts as a subsonic piston and drives a large-amplitude simple wave which can then shock and produce the observed global coronal \corr{wave}. If the perturbation is sufficiently strong \corr{in the low corona at heights $\leq100$~Mm} (in this case with \corr{both} a Mach number \corr{$M_{A}$ and density compression ratio $X = n_{1}/n_{2}$ of $\approx 2$}), the downward propagation of the plasma flow induces a chromospheric signature observed as the Moreton--Ramsey wave. As shown in Figure~\ref{fig:int_vary}, the density compression ratios and Mach numbers estimated for each of the events presented here are much lower than those predicted by \citet{Vrsnak:2016}, yet in each case a Moreton--Ramsey wave is observed. Given the inevitable complications when comparing idealised simulations with \corr{real observations}, there are several possible solutions for this discrepancy which require discussion.

The first issue here is that the global waves observed here originate in an active region located at $\approx$N10$^{\circ}$ and propagate towards the north solar pole. As a result, there are increased foreshortening effects which must be accounted for when considering each event. The estimated density compression ratio and Mach numbers should therefore be considered to be a minimum estimate of the true values. While it is true that the numbers given here are at the lower end of previous observations \citep[for example, those presented by][]{Muhr:2011,Long:2015}, all of these previously observed density compression ratios are much lower than those predicted by \citet{Vrsnak:2016} from simulations, suggesting that this explanation is insufficient.

An alternative explanation is that in addition to the lateral expansion of the global \corr{waves} observed here, they had a much stronger downward component than previously observed events \citep[\cf][]{Warmuth:2004b}. As noted by \citet{Vrsnak:2016}, it would be possible for a highly asymmetric but weakly impulsive eruption to produce a Moreton--Ramsey wave on the erupting side of the flux rope. This is consistent with the observations shown in Figure~\ref{fig:context}, where each of the global wave events can be seen to be highly asymmetric, propagating primarily towards solar north. An examination of the large-scale magnetic field in the proximity of the erupting active region \citep[obtained using the PFSS method described by][]{Schrijver:2003} and the resulting magnetic pressure shown in panel c of Figure~\ref{fig:b_field} shows that while the magnetic pressure is quite large above and south of the erupting active region, it drops off significantly moving from the active region towards the north pole (\ie, in the direction of propagation of the global \corr{waves}). This suggests that each of the eruptions could have been highly inclined as a result of following the path of least resistance, as suggested by \citet{Panasenco:2013}  \citep[see also the eruption from 8-April-2010 modelled by][]{Kliem:2013}. 

As a particularly well observed eruption event, the X1 flare on 29-March-2014 has been studied by multiple authors (see Section~\ref{s:intro}). This analysis has included several magnetic field extrapolations \citep[\cf][]{Woods:2018} and spectroscopic analysis of the plasma evolution within the erupting filament before and during the flare \citep[\cf][]{Kleint:2015,Woods:2017}. As noted in each case, for this event the eruption occurred on the northern side of the active region, and was primarily driven by flux cancellation in the middle of the active region. Global EUV waves have previously been observed to propagate asymmetrically when originating on the edges of active regions due to the increased Alfv\'{e}n speed inhibiting propagation through the active region \citep[\cf][]{Long:2008}. This is again consistent with the increased magnetic pressure above and to the south of the erupting active region shown in Figure~\ref{fig:b_field}.

In addition to this, the NLFFF extrapolation in panels a and b of Figure~\ref{fig:b_field} \citep{Valori:2010} confirms, first, the presence of a flux rope in the northern edge of the active region, and, second, the presence of a complex strong magnetic structure to the south of the origin of the eruption which could also have contributed to the strongly anisotropic nature of the global wave. The large structures (seen in red and blue in panels a and b of Figure~\ref{fig:b_field}), \corr{would have inhibited the eruption and early evolution of the flux rope, forcing it to erupt asymmetrically towards solar north. This asymmetric eruption would have driven a global EUV wave with a significantly increased downward component which would have pressed down on the chromosphere and been observed as a Moreton--Ramsey wave.} The blue \corr{magnetic field} structure also expands out above the erupting flux rope (seen as the small red structure to the top and right of panels a and b in Figure~\ref{fig:b_field} respectively), \corr{which} is compatible with the (current-free) large-scale structure found in the PFSS extrapolation of panel c. Together, the two extrapolations support the \corr{conclusion} that the ambient field could have contributed to the increased downward component of the global waves studied here. \corr{While the multiple eruptions would have been expected to change the topology of the surrounding magnetic field, this would be on a local rather than global scale and involves removing the currents associated with the erupting flux rope. As the PFSS is current-free, the eruptions do not affect the magnetic pressure estimated here and as a result it remains comparable to that shown in Figure~\ref{fig:b_field} over the timescale discussed here.}

\section{Conclusions} \label{s:conc}

The series of eruptions associated with AR~12017 provide a unique opportunity to study the evolution of 4 homologous global waves in the solar atmosphere using both EUV and H-alpha observations. Despite the long history of observations of these features, joint observations of global EUV and Moreton--Ramsey waves, particularly with very high spatial and temporal cadence, continue to be incredibly rare. Although Moreton--Ramsey waves are known to be the chromospheric footprint of a global wave propagating in the solar corona \citep[\cf][]{Uchida:1968,Warmuth:2015}, far fewer observations have been made of Moreton--Ramsey waves than the regularly observed global EUV waves. However, simulations performed by \citet{Vrsnak:2016} suggest that the global EUV wave must be sufficiently strong for it to produce an observable perturbation of the high density chromosphere. 

The events originating from AR~12017 from 28--30 March 2014 provide an opportunity to test this hypothesis. All four homologous waves were well observed in the corona by SDO/AIA and in the chromosphere by GONG, enabling a direct comparison between the events and the different regimes of the solar atmosphere. The wave kinematics were measured using multiple passbands, with the global EUV waves exhibiting high velocities and strong decelerations, consistent with previous results. The Moreton--Ramsey waves were found to have lower velocities and weaker deceleration, consistent with both the known picture of how Moreton--Ramsey waves are produced \citep[\cf][]{Warmuth:2004b} and the lower cadence of the H-alpha observations \citep[\cf][]{Byrne:2013}. 

With the kinematics of each event indicating that the global waves were weakly shocked, the density compression ratio was estimated in each case using both an intensity ratio and DEM approach. The Mach number could then be estimated for each event, with all four waves found to be very weakly shocked with Mach numbers of 1.02--1.20. Although consistent with previous work \citep[\cf][]{Long:2015}, these Mach numbers were below the numbers predicted from simulations by \citet{Vrsnak:2016}, suggesting that none of the global waves should have produced a Moreton--Ramsey wave.

The magnetic structure of the erupting active region was examined to determine an alternative explanation for how each event could therefore perturb the chromosphere and produce a Moreton--Ramsey wave. It was found that each of the eruptions occurred to the north of the source active region, with the erupting flux ropes found to originate underneath an expanded magnetic loop structure. Global waves have been observed to be produced by a rapid overexpansion of the erupting flux rope, which acts as a piston and drives a shock front which then propagates freely \citep[\eg][]{Patsourakos:2010}. However, the structure of the surrounding magnetic field found here suggests that instead of being driven laterally across the solar disk as with other events, the global wave had a significant downward component \corr{as a result of the significant asymmetric eruption of the flux rope. This therefore enabled the global EUV wave} to perturb the chromosphere and produce a Moreton--Ramsey wave, despite being weaker than the limit predicted by simulations.

\acknowledgments
The authors wish to thank Julia Lawless for useful discussions \corr{and the anonymous referee for their very detailed suggestions which helped clarify the manuscript}. DML acknowledges support from the European Commission's H2020 Programme under the following Grant Agreements: GREST (no.~653982) and Pre-EST (no.~739500) as well as support from the Leverhulme Trust for an Early-Career Fellowship (ECF-2014-792) and is grateful to the Science Technology and Facilities Council for the award of an Ernest Rutherford Fellowship (ST/R003246/1). GV was funded by the Leverhulme Trust under grant 2014-051. JMJ thanks the STFC for support via funding given in his PhD Studentship.

\vspace{5mm}
\facilities{SDO, GONG}
\software{SolarSoftWare \citep{Freeland:1998}}

\end{document}